\documentclass[conference]{IEEEtran}

\ifCLASSINFOpdf
\else
\fi
\usepackage[]{graphicx}
\usepackage{tikz}
\usepackage{url,balance}
\usepackage{enumitem,color}
\usepackage{colortbl}
\definecolor{Gray}{gray}{0.9}
\def\etal{\emph{et al. }}
\usepackage[normalem]{ulem}

\begin{document}
%

\title{How to Make Privacy Policies both GDPR-Compliant \emph{and} Usable}

\author{
\IEEEauthorblockN{Karen Renaud}
\IEEEauthorblockA{School of Design and Informatics\\
Abertay University\\
Dundee, United Kingdom\\
k.renaud@abertay.ac.uk}
\and
\IEEEauthorblockN{Lynsay A. Shepherd}
\IEEEauthorblockA{School of Design and Informatics\\
Abertay University\\
Dundee, United Kingdom\\
lynsay.shepherd@abertay.ac.uk}}

\maketitle

\IEEEpeerreviewmaketitle

\begin{abstract}
It is important for organisations to ensure that their  privacy policies  are General Data Protection Regulation (GDPR) compliant, and this has to be done by the May 2018 deadline.  However, it is also important for these policies to be designed with the needs of the human recipient in mind. We carried out an investigation to find out how best to achieve this. 

We commenced by synthesising the GDPR requirements into a checklist-type format. We then
 derived a list of usability design guidelines for privacy notifications  from the research literature.
We augmented the recommendations with other findings reported in the research literature, in order to confirm the guidelines.
We  conclude by providing a usable and
 GDPR-compliant   privacy  policy template for the benefit of policy writers. 
\end{abstract}


\section{Introduction}
Those who surf the web risk having their privacy violated. They need to be informed about what personal data websites are collecting so that they can choose to patronise those who do not violate their privacy, or opt out of the use of their information. In other contexts there is evidence that people do respond to warnings \cite{cox1997product,silic2015warning},  
with confirmation from a study in the privacy context \cite{egelman2008you}.
Yet it is non-trivial to design effective  privacy policies 
\cite{larose2007promoting}. 

Obar and Oeldorf-Hirsch \cite{obar2016biggest} found that 74\% of the 543 people in their study did not even read the privacy policy. Where websites force 
users to read and agree to their policies (e.g. Google), 
they often become discouraged and overwhelmed  because  the  text is overly long or incomprehensible \cite{reidenberg2015disagreeable}. Computer users often receive  too many privacy advisements \cite{egelman2008you,kim2009habituation,anderson2014users}, and  
sometimes do not know what actions to take as a consequence of policy information \cite{shankar2006doppelganger}.

The Web Content Accessibility Guidelines{\footnote{\url{https://www.w3.org/TR/WCAG21/}}} require notifications to be  perceivable, operable, understandable and  robust \cite{almeidamerging}. 
The evidence from investigations into privacy policy examples suggests that they do not demonstrate these qualities \cite{reidenberg2015disagreeable}.
This diminishes the efficacy of policy notifications, and leaves users vulnerable to unknowingly carrying out actions that will compromise their privacy.

The advent of GDPR adds another level of complexity to the design of privacy policies.   
Guidance provided by the Information Commissioner's Office \cite{GDPRPrep} stresses the importance of communicating the necessary privacy information to stakeholders, and raising awareness as to the impact of how the organisation implements GDPR requirements. GDPR also requires 
privacy policies  to  deliver their message  effectively, efficiently and to the user's satisfaction i.e. usably.

Usability methods seek to make policies look less like legal documents, usually worded in legalese,
ensuring that the man and woman in the street is able to understand it.
Legalese prevents computer users being given fair notice due to poor understandability \cite{Waldman16}.
Clear and unambiguous communication is, in essence,  the \emph{raison d\^{e}tre} of privacy policies.

The problem is perhaps that traditional usability guidelines cannot necessarily be used ``as-is'' in the  privacy context because usability testing is usually related to primary task completion. Privacy, on the other hand, is seldom the end user's primary task \cite{balebako2013little,kelley2009designing}. That being so, the display of a privacy policies can interrupt the user's pursuit of their primary goal and is thus often perceived to be a nuisance  \cite{albrechtsen2007qualitative}.
We  need bespoke guidelines to inform policy design in the  privacy context.

Waldman \cite[p. 8]{Waldman16} reports that their review of 191 privacy policies convinced them that ``{\em today's privacy policies are not designed with readability, comprehension, and access in mind}''. This justifies the need for explicit usability guidelines to be provided to web privacy policy writers.

Our work  seeks to inform policy writers, with guidance that is  specifically tailored towards  browser-based  privacy policies that are both usable and GDPR-compliant. 
 
We first detail the context of our investigations in Section \ref{preamble} then summarise the GDPR legislation requirements in Section \ref{GDPR}. We  carried out a systematic literature review of design guidelines for designing usable privacy policies (Section \ref{analysis}). 

To make our guidelines as helpful as possible, we decided to convey the {\em spirit} of the guidelines in the form of a privacy policy template. This conveys  the ``how'' of privacy policy design, rather than the ``what'', as encapsulated in a linear set of policy design guidelines.
The paper provides a template pattern for a policy that is both usable and GDPR compliant (Section \ref{finaltemplate}), before concluding in Sections \ref{future} and \ref{conc}

\section{Preamble}\label{preamble}

Wogalter and Mayhorn \cite{Wogalter2017} explain that warnings (policy items) are a type of risk communication. Wogalter \cite{wogalter1999factors} explains that warnings have two purposes, to: (1) communicate information, and (2) reduce unwise behaviours. To achieve these aims, the policies have to be designed carefully.

To understand how humans process communications, we need to look at how researchers have modeled this. 

Wogalter, DeJoy, and Laughery \cite{wogalter1999organizing} developed the C-HIP model in the context of warning research. Their model builds on initial human communication models proposed by Shannon \cite{shannon2001mathematical} and Lasswell \cite{lasswell1948structure}.
Wogalter {\em et al.}'s model can be considered to be somewhat unrealistic because it does not include a noise component, as Shannon's does. In a world of noisy communication such a model can not be complete. Cranor \cite{cranor2008framework} proposed a human-in-the-loop framework which is more comprehensive and reflects the factors impacting communications in the  context of \emph{security}  notifications.

It is important to realise that security and privacy are fundamentally different concepts. 
Skinner \etal \cite{skinner2005framework} argue that a secure information system does not necessarily imply that privacy will be preserved in the system. 
 Gritzalis and Lambrinoudakis \cite{gritzalis2008privacy}, and Bambauer \cite[p. 667]{bambauer2013privacy} make similar arguments.
 As an example, they refer to a company that collects customer information and stores it in an encrypted format. This ensures that the information is secured. Yet the same company may sell the information to another company, thereby violating the owner's privacy.

 Privacy and security, being clearly distinct concepts, require different models of notification design. This means that we cannot merely use the security communication processing models to inform the design of privacy policies. In the absence of a published privacy communication model, we plan to use the GDPR legislation to structure our privacy policy design guidelines.

\section{GDPR Legislation}\label{GDPR}
The introduction of the GDPR is said to be \textit{``the most important change in data privacy regulation in 20 years"} \cite{GDPRHome}.  The legislation will come into force on the 25th May 2018, and replaces the existing Data Protection Directive 95/46/EC. Organisations that fail to comply will be subject to significant fines. The main GDPR requirements are that customers must be informed about (numbering is ours):

\subsubsection*{\textbf{GDPR1: Specify Data Being Collected}}
Customers should be aware of the information that is collected about them.  Furthermore, businesses should document the information that is collected, which links into the accountability required by GDPR \cite{GDPRPrep}.

\subsubsection*{\textbf{GDPR2: Justification For Data Collection}}
Organisations must explain their rights to collect data \cite{GDPRPrep}, but they should also justify to themselves exactly \textit{why} they need to collect such information \cite{GDPRJust}.

\subsubsection*{\textbf{GDPR3: How Data Will Be Processed}}
The organisation must inform the customer of the lawful rights it has to process personal data \cite{GDPRPrep}.

GDPR outlines the ways in which processing is deemed legal (one of the following must apply): the customer has given consent for this to be done for a specific purpose, it is used to form a contract with a customer, the data controller is complying with a legal obligation, it is used to protect the interests of a person, it is required for a task involving the public interest, it is required for a legitimate purpose by the controller (provided rights and freedoms are not violated) \cite{GDPRArt16}. Moreover, the person has the right to opt out of processing of his data by an algorithm, or any other profiling. 

Under Article 9 of the legislation, there is a special category of data, deemed `sensitive data' which requires further protection.  This information can include details of an individual's health, political views, religion, etc.
A lawful basis for processing such information must be given (these have been outlined in a previous paragraph), and a separate basis must be provided for processing special category data \cite{GDPRArt9}.
Examples of reasons for processing such data include: it may be necessary for reasons of public health, or it may be necessary for the progression of legal claims \cite{GDPRArt9}.
\subsubsection*{\textbf{GDPR4: How Long Data Will Be Retained}}
GDPR dictates that data should be held for the minimum amount of time, and organisations must state how long data is retained \cite{GDPRPrep} \cite{GDPRDataRet}.

\subsubsection*{\textbf{GDPR5: Who Can Be Contacted to Have Data Removed or Produced}}
People have the right for all their data, both provided to the company, and observed by their systems: (1) to be forgotten, and (2) to be provided to them.
To facilitate this, contact details must be provided in the policy \cite{durkan2003exploring,gantner2015all}.
Within the organisation, someone must take responsibility for the  stored and processed data.  Customers should also be informed who the Data Protection Officer (controller) is, and how to  get in touch with them, should they have an access request \cite{GDPRPrep}.
Customers should also be provided with a timescale in terms of how subject access requests will be handled by the organisation \cite{GDPRPrep}.

\subsubsection*{\textbf{GDPR6: Communication of Privacy Information}}
Documentation on the legislation notes that it ``{\em requires the information to be provided in concise, easy to understand and clear language}'' \cite{GDPRPrep}.

\begin{figure}[ht]
	\centering
        \includegraphics[width=\columnwidth]{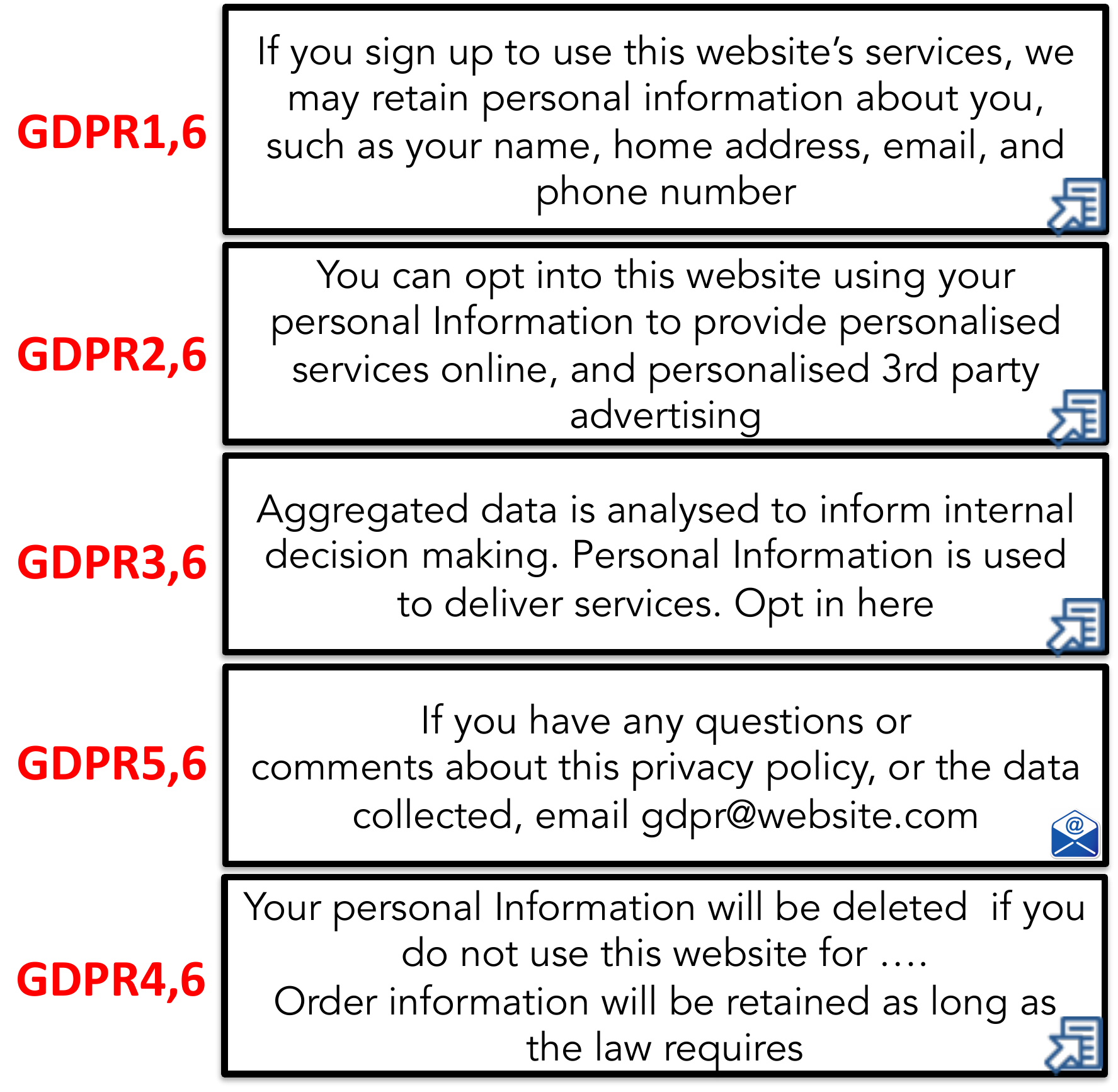}
	\caption{GDPR-Compliant Policy Template. Each section provides a link to more comprehensive information}
	\label{fig:template}
\end{figure}

We now present a
GDPR-compliant policy template in Figure \ref{fig:template}.

\subsection{Assessing Current State of Play}
To take a snapshot of the current situation, roughly three months before the GDPR deadline, we proceeded to assess the privacy policies of some UK-based websites. We carried out this assessment on the 25th January 2018.

In order to choose the  UK websites to assess, we
consulted  Alexa  to obtain the top 10 most-used websites in the UK{\footnote{\url{https://www.alexa.com/topsites/countries/GB} 
Alexa uses web traffic analysis to produce lists of the most popular websites in countries worldwide}}. 

The {\bf first} step is to be able easily to locate the policy.
Langhorne \cite{langhorne2014web} reported, in 2014, that many organisations did not provide a handy link to their privacy policies from the landing page. It is likely that the upcoming GDPR legislation will mandate provision of such links. 
All of the websites we examined did indeed include a link to their privacy policy from their main page, which was a positive development.

{\bf Secondly}, we checked the extent to which the websites' privacy policies 
satisfied GDPR requirements. To provide a measure of understanding (GDPR6), we  use the Gunning Fog Index score{\footnote{\url{http://gunning-fog-index.com/}}}.
This index is an indication of the number of years of schooling someone would need to be able to understand the text. If someone needs more than a high school education to understand the policy (more than 13 years), we conclude that it fails GDPR6 in terms of understandability. Table \ref{websites} presents our findings. 

\begin{figure}[ht]
	\centering
        \includegraphics[width=\columnwidth]{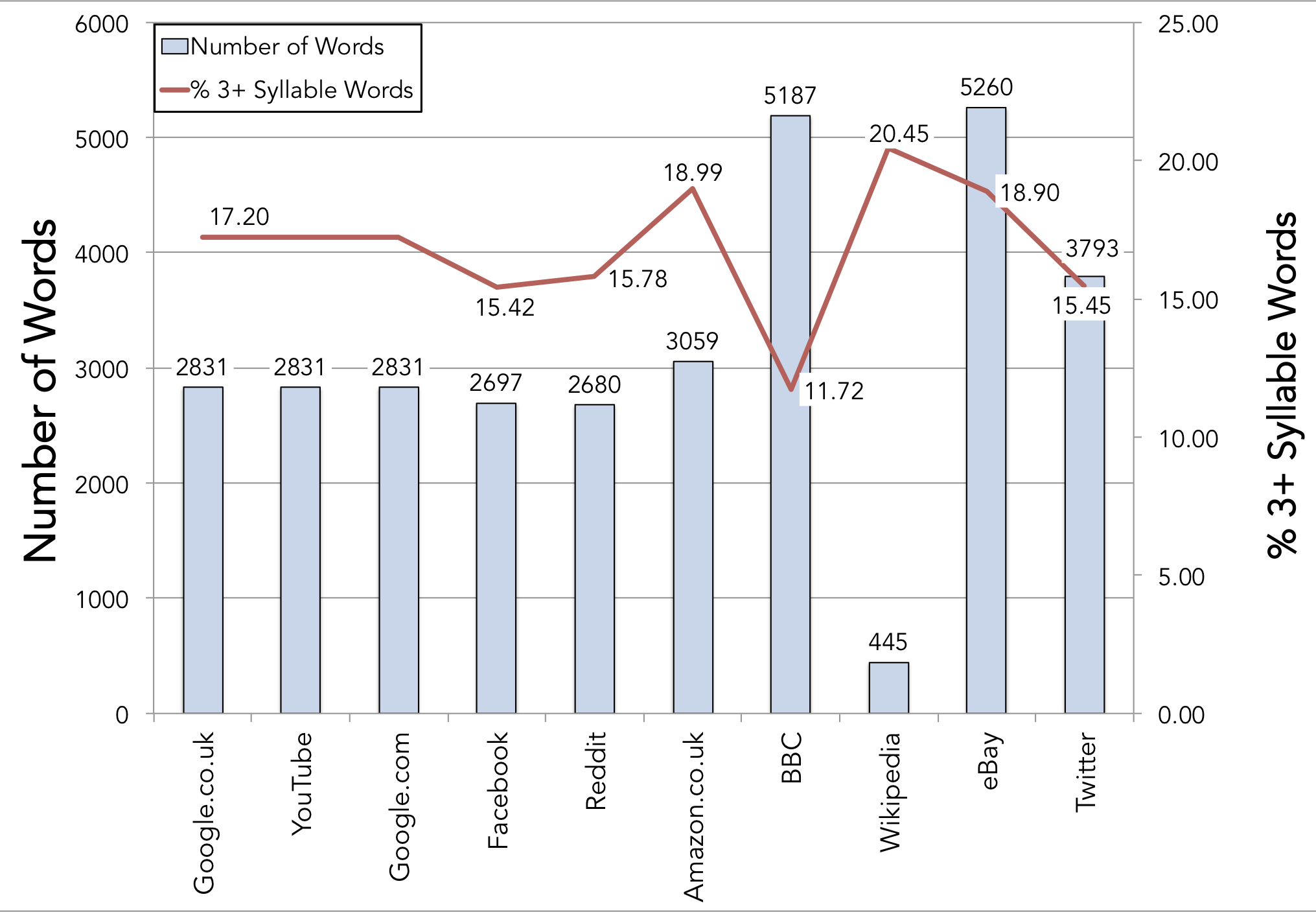}
	\caption{Word Lengths and \% of Complicated Words (3+ syllables)}
	\label{fig:gr}
\end{figure}

We also provide the number of words in total, as well as the number of complicated words (with 3 or more syllables) to give an idea of the effort a user would have to expend if they wanted to read and understand the entire  policy. 
The  data is depicted in Figure \ref{fig:gr}.

\begin{table}[ht]
\begin{tabular}{|p{1.5cm}|c|c|c|c|c|c|p{1cm}|p{1cm}|p{1cm}}
\hline
  GDPR Number   &  1 & 2 & 3 & 4 & 5 & \multicolumn{3}{c|}{6}\\
  \cline{7-9}
  &&&&&&GFI&Words&3+ Syllable Words\\
  \hline
  Google.co.uk   & $\bullet$ & $\bullet$ & $\bullet$ &  	$\otimes$ &$\otimes$ & 15.21&2831&487\\
   YouTube   & $\bullet$ & $\bullet$ & $\bullet$ &  $\otimes$ &$\otimes$ & &&\\
Google.com   & $\bullet$ & $\bullet$ & $\bullet$ &  $\otimes$ &$\otimes$ & &&\\
\rowcolor{Gray}
  Facebook & $\bullet$ & $\bullet$ & $\bullet$ &  $\otimes$ & $\otimes$& 13.71&2697&416\\
  Reddit & $\bullet$ & $\bullet$ & $\bullet$ &  $\otimes$ & $\otimes$ & 13.86&2680&423\\
 \rowcolor{Gray}
   Amazon.co.uk & $\bullet$ & $\bullet$ & $\bullet$ &  $\otimes$ & $\bullet$ & 12.21&3059&581\\

  BBC$^*$ & $\bullet$ & $\bullet$ & $\bullet$ &  $\bullet$ & $\bullet$ &  11.34&5187&608\\   
\rowcolor{Gray}
  Wikipedia & $\bullet$ &$\bullet$ & $\bullet$ &  $\bullet$ & $\otimes$ & 13.74&445&91\\

  eBay & $\bullet$ & $\bullet$ & $\bullet$ &  $\otimes$ & $\otimes$ & {\bf 17.97}&5260&994\\
\rowcolor{Gray}  
 Twitter & $\bullet$ & $\bullet$ & $\bullet$ &  $\otimes$ & $\otimes$ & 13.51&3793&586\\
\hline   

\end{tabular}
	\caption{Top Alexa Websites and GDPR Requirements. Starred Websites are GDPR Compliant.\newline (GFI=Gunning Fog Index: $\bullet$=satisfies; $\otimes$=does not satisfy)}
	\label{websites}
\end{table}

Only one of these policies met the requirements of the GDPR legislation on the 28th January 2018. There is still time left for the others to revise their policies and they will probably do so, most being large companies with substantial web development resources at their disposal. Yet smaller companies would probably benefit from some guidance in this respect.

In the next section we consider what the research literature says about how to design privacy policies. 

\section{Usability Guidelines}
\label{analysis}

We decided to focus on browser privacy policies firstly because of the popularity of web applications  \cite{mikowski2013single} such as email, claimed to be the most popular application in use \cite{alharbi2008graphical}. Video streaming \cite{kellerman2010mobile}, which runs within browsers, is also very popular.
The second reason is that browsers run on all devices, ranging from Desktops to Smartphones. We thus felt that our guidelines could be most useful to developers if we focused on guidelines for browser application policy writers.

We carried out a systematic literature review in order to gather best practice from the research literature in this respect.

\subsection{Systematic Literature Review }

The literature search was carried out in January 2018 as follows: 

{\bf Databases:} ACM, Springer, Web of Science, Scopus, IEEE, and then Google Scholar to identify publications that did not appear in the databases.

{\bf Keywords:} `design guidelines' {\em and} `browser'{\em and} `privacy' {\em and} (`feedback' {\em or} `warnings' {\em or} `notification' {\em or} `alert').  A separate search was conducted using the phrase `privacy policy design'.

{\bf Time Range:} 2007---2017

 {\bf Exclusion Criteria:} Patents, citations, non-peer reviewed, not English or unobtainable.

\begin{table}[ht]
\begin{tabular}{l|>{\hfill}p{1.5cm}|>{\hfill}p{1.5cm}|>{\hfill}p{1.5cm}}
  Database   & Returned &  Excluded &  Analysed \\ \hline
  Scopus   & 0 & 0 & 0\\
\rowcolor{Gray}
  ACM & 3 & 2 & 1\\

  Springer & 145 & 139 & 6\\
\rowcolor{Gray}
  Web of Science & 0 & 0 & 0\\

  Google Scholar & 61 & 42 & 19\\
\rowcolor{Gray}
  IEEE & 73 & 70 &  3\\
  \hline
  Total & \multicolumn{3}{r}{29}\\

\end{tabular}

	\caption{Papers from the literature search}
	\label{litsearch}
\end{table}

We analysed the guidelines using Thematic Analysis \cite{guest2012introduction}. This approach supports pinpointing, examining, and recording themes that emerge from the papers. We commenced by familiarising ourselves with the papers. We then generated initial codes and searched for themes as we collated these codes. We then reviewed the themes, defining and naming them. 
Finally, we assigned them to the applicable GDPR category, as detailed in Section \ref{GDPR}.

   \subsection{Results}
   
 \subsubsection*{\textbf{GDPR1}}
 Ensure that the sensitivity of the data is communicated to the user \cite{nafra14}. This need is confirmed by \cite{liu2005beyond}.
 
 \subsubsection*{\textbf{GDPR2}}

Some researchers advise that providing justifications for privacy policies potentially reduces the end-user's trust in the system \cite{aagaard2013privacy,adjerid2013sleights,Knijnenburg15,pollach2007s}. 
Volkamer \etal \cite{DBLP:conf/trust/VolkamerRCRB15} advise that the potential consequences of a  risk  be conveyed to the user, along with potential recommendations.
GDPR  mandates that this information be provided so we should focus on fostering trust in the presence of such justifications.

  \subsubsection*{\textbf{GDPR3}} ---
  
   \subsubsection*{\textbf{GDPR4}} ---
   
    \subsubsection*{\textbf{GDPR5}}
    
    It is important to ensure that the user can contact someone in the organisation \cite{goldberg2009state,GhazinourA16}. Contact details should be conspicuously placed   \cite{PRIME}.
    
     \subsubsection*{\textbf{GDPR6}}
In this section we first present the themes that emerged from our analysis. We then cite supporting research from other publications. The themes fell naturally into two meta categories: (1) content of the policies, and (2) delivery of the policies. We report these separately\\
\ \\ 
\textbf{\large\emph{Content Guidelines:}}\\ 
 The overarching admonition should be that human attention is a finite resource
\cite{colnago2016privacy,GhazinourA16} that should not be taken for granted or squandered, and privacy policies \textit{``should empower users to make informed decisions about their online behavior''} \cite{DBLP:conf/hci/Langhorne14}.\\
\ \\
  \indent  \textbf{\emph{(a) Modality ---}} Murphy-Hill \& Murphy \cite{murphy2014recommendation,Westermann17} suggest that pictures be used to ease communication because users prefer this \cite{Chen:2011:IDE:2008579.2008601}.  Others have advocated visualising privacy policy statements, making them more usable \cite{GhazinourA16}.

    On the other hand, Goldberg \cite{goldberg2009state} suggests that text should be used exclusively to maximise accessibility. 
    Anderson \etal \cite{anderson2015polymorphic, anderson2016warning} suggests the use of polymorphism in warning notifications to reduce habituation.

 \emph{\underline{Supporting Research:}} Other researchers argue for the power of a multi-modality image and text message in enhancing communication \cite{karimov2011effect,Merchant,messaris1997visual,messaris2001role}.

\textbf{\emph{(b) Make it Personal ---}}
Vasalou \cite{vasalou2015understanding} says policy items should give recipients ``space for interpretation'', so that they can understand how it applies particularly to them \cite{lin2013understanding}. 

The personalisation of policies should be considered  \cite{Chen:2011:IDE:2008579.2008601, nursespi2013, nurse2011past,redmiles2017}.
.

\emph{\underline{Supporting Research:}} Elman \etal \cite{elman2000aphasia} argue that personalisation, by whatever means, is extremely important in enhancing understanding. 
Schaub \etal \cite{schaub2017designing} says privacy policy notices should be ``relevant'' to the person. 
Needham \cite{needham2011personalising} also argues for the importance of personalisation. Yet policy display is somewhat different from other kinds of personalisation opportunities: people view the policy {\em before} they have divulged any information that could be used to personalise the communication. 
That being so,
one way of personalising a generic document, such as a policy, especially in helping people to see that it applies to them,  could be by using personal pronouns like ``you'' and ``your''. This should help people to consider the personal ramifications of the policy.

Another way is to provide examples that people can identify with \cite{robinson2010personalising}, but this will take up valuable space and needs detailed investigation to assess viability.

\textbf{\emph{(c) Give Control to the User ---} }
 It is important for the user to retain a level of control \cite{PRIME, vasalou2015understanding,xu2012value} by allowing them to exercise control over disclosure \cite{colnago2016privacy}. 
 Schaub \etal \cite{schaub2015design} distinguish between three levels  of user control: (1) blocking, non-blocking and decoupled. 
 A designer has to
 decide whether the user has to acknowledge the policy notification (blocking) or not (non-blocking), whether they can defer their response (decoupled), or whether the option's actions will expire  \cite{murphy2014recommendation}.
 
 Users should be provided with the option to respond to a  risk they have been notified about, and helped to visualise potential consequences \cite{nursespi2013,jonesprobing}. 
 
  \emph{\underline{Supporting Research:}} Other research emphasises the need to allow people to control disclosure \cite{liu2005beyond,schaub2017designing}. Yet Waldman \cite{Waldman16} reports that, of the 191 policies they surveyed in 2016, only 9 provided users with noticeable opt-out buttons. Moreover, they discovered that  a little more than half of these only allowed users to opt out of marketing, but not out of profiling. GDPR mandates that users should be allow to opt out of the latter. Yet Adjerid \etal \cite{adjerid2014framing} point out that merely allowing people to opt out, without carefully considering the way the information about such consent is presented to the user is framed, does not necessarily help them to make better privacy choices.

\textbf{\emph{(d) Trust ---} }Trust should be deliberately built and maintained \cite{murphy2014recommendation,Waldman16} by framing the privacy policy very carefully \cite{adjerid2013sleights}. Indeed, when people read privacy policies, it impacts on their trust of the website \cite{martin2015formal}, so it is important to get it right.

It is crucial for people to trust a website if they are to make use of it \cite{sun2014understanding}. Broutsou and Fitsilis \cite{broutsou2012online}
 review the literature on trust and report a number of studies that show
  that the level of trust is  positively related to the intention to carry out an online  transaction.
 
   \emph{\underline{Supporting Research:}}  
 Other research suggests that users require reassurance that information is kept securely  \cite{liu2005beyond,GhazinourA16} and recommend including a Privacy Seal  \cite{johnston2003security,durkan2003exploring,gantner2015all,sun2014understanding}.
Policy writers should also provide a telephone number (not only an email address) and make other channels of communication clear \cite{durkan2003exploring,gantner2015all}. 
Finally, the policy should explain how these privacy assurances will be enforced \cite{wu2012effect,doty2013privacy}.

\textbf{\emph{(e) Overview \& Link ---} }
 Lin \cite{lin2013understanding} suggests highlighting the most important information. We should only present essential details about the risk  \cite{nursespi2013, nurse2011past}, with links to more information should they want it \cite{DBLP:conf/trust/VolkamerRCRB15}.  In providing  policy-based notifications, a balance must be found between brevity and comprehensiveness \cite{balebako2013little}.
 
 {\emph{\underline{Supporting Research:}}} Researchers confirm the need to provide an overview first and then links to more information  \cite{gantner2015all,johnston2003security}

\textbf{\emph{(f) Maximise Understandability ---} } This is emphasised by a number of researchers \cite{kelley2009designing,murphy2014recommendation,Westermann17,shah2016evaluating,jonesprobing,kununka2017end,Waldman16} as well as  the importance of consistency \cite{almeidamerging,murphy2014recommendation}.

Unclear notifications are more likely to be ignored, and consideration should be given to the exact meanings of words used \cite{nursespi2013}.
Concrete explanations should be provided \cite{ozok2010design,Ng2017} and  explanations should be simple \cite{lin2013understanding}, avoiding
acronyms and jargon with only meaningful terminology being used \cite{balebako2013little,kelley2009designing,shah2016evaluating,silic2015warning}.  Semantically distinct information should be separated \cite{kelley2009designing,vasalou2015understanding}. 
Text  should be presented in  short, simple sentences, devoid of complex grammatical structures \cite{harbach2013sorry, Harbach:2012:TMW:2382196.2382301, Pala:2009:UUI:1927830.1927852, DBLP:journals/telsys/DongCJ10}. Longer warning notifications performed poorly in user testing \cite{DBLP:conf/interact/Bravo-LilloCDKS11}.

Some users may have low numeracy levels so that other mechanisms for communicating risk should be sought. 
In choosing these, it should be borne in mind that users may have different understanding of visuals  \cite{nursespi2013}. 
 
    {\emph{\underline{Supporting Research:}}}  
    Other authors confirm the importance of maximising ease of use \cite{sun2014understanding,Hertefelt,schaub2017designing}.
    
In terms of understandability, it must be noted that  existing work confirms that shorter notifications are most effective at communicating with users. The challenge, in providing enough information to foster  understanding, while being brief, is highlighted \cite{felt43265}. 

\ \\
\ \\
\textbf{\large\emph{Delivery Guidelines:}}\\   
\indent  {\emph{(i) Timing \& Location ---}} 
Many of the recommendations that fall into this category are related to the delivery of pop-up type alerts and notifications, both in terms of time and space. There is a focus on displaying these 
 only when they merit  interrupting the user's task \cite{Westermann17,akhawe2013alice},  avoid irritating  \cite{DBLP:conf/trust/VolkamerRCRB15} and  preventing habituation \cite{murphy2014recommendation,akhawe2013alice}. 
     
 Privacy policies, unlike these kinds of alerts, are  either viewed when the person deliberately clicks on a link, or is forced to read the policy and consent to it.  Hence time and space are less applicable in this context.

{\emph{(iii) Appearance ---}}  Kelley \cite{kelley2009designing} provides a number of recommendations: (1) the notification should be surrounded by a box to clearly demarcate it;
 (2)  provide a title to assist speedy recognition.
It is important to  
     be careful with colour use so as not to disadvantage those with colour deficiencies \cite{goldberg2009state}.
A neutral grey colour can be used for the background of notifications, as it is unlikely to annoy the user \cite{DBLP:conf/trust/VolkamerRCRB15}.

\subsection{Reprise}
It is clear from the previous discussion that much attention has been given to guidelines to ensure that GDPR6 is satisfied.  GDPR3 and GDPR4  requirements were not addressed in the literature we gathered, while GDPR1 and GDPR5 did not receive much attention. GDPR2 is an area ripe for focused attention, because many of the current guidelines conflict with the GDPR requirements.

We could simply provide the list of content-related guidelines based on the derived principles in the previous section.
However
designers have difficulty benefiting from these kinds of flat lists of guidelines \cite{Renaud17,luger2014value}. We therefore plan to produce a template to demonstrate the impact of these guidelines. Waldman \cite{Waldman16} discovered that a demarcated structure for policies made them more palatable to users.

 \section{Usable and GDPR-Compliant Privacy Policy Template}
 \label{finaltemplate}

 In this section we consider how to implement the content guidelines from the literature as described in the previous section. 
 The delivery guidelines will not be considered because they have a great deal to do with the context and nature of the website and cannot be provided in a context-neutral fashion.

In providing an example GDPR-compliant template, we  formulated text to deliver the content for a fictional Company X, as advised by the GDPR requirements and content guidelines. We measured the understadability of the text by using the Gunning Fog Index test. 
 
Some of the content guidelines are relatively easy to satisfy, more or less in a binary fashion i.e. overview \& link. Guidelines (d) (trust) and (f) understandability, require a more nuanced approach. 

\textbf{\emph{GDPR6(d) Trust:}} To address trust issues we decided to include an image to foster and inspire trust. We decided to propose the use of a Privacy Seal for this purpose, especially since this has been widely advised \cite{johnston2003security,durkan2003exploring,gantner2015all,sun2014understanding}. Moreover, we shall include icons in each subsection to demarcate them and improve accessibility.

\textbf{\emph{GDPR6(f) Maximise Understandability:}} To maximise the understandability required by GDPR6, we simplified the text to need less than a high school education to understand, and included a small icon to bookmark different sections.

The years of compulsory schooling a person receives depends on the country they are from.  For example, in the UK, children attend school from the ages of 5 to 18, however they are free to leave at the age of 16, meaning they can receive between 11 and 13 years of schooling.  In contrast, when considering other EU countries, Swedish children start school at the age of 7, and can leave at 16, meaning they may only receive 9 years of schooling.

Research presented in this paper was conducted by an English-speaking, UK-based institution, therefore the assumption was made that people typically have between 11 and 13 years of schooling.
 Table \ref{tab:TemplateGFI} provides the GFI of the text provided to address all the GDPR requirements as understandably as possible.

\begin{table}[ht]
\centering

\begin{tabular}{p{1cm}|p{0.5cm}|p{5cm}}

Guideline & GFI   & Text Used     \\ \hline
GDPR1     & 8.457  & If \textbf{you} sign up to use this website's services, we may keep personal information about \textbf{you}. This will include \textbf{your} name, home address, email, and phone number \\  \rowcolor{Gray} 
GDPR2, GDPR6(c)    & 10.30 & This website will use \textbf{your} information  to provide better services to \textbf{you}, and adverts from 3rd parties.  Opt out here \\ 

GDPR3     & 5.822 & We would like to collect all order information to help us to predict global trends. Opt in here
  \\  \rowcolor{Gray} 

GDPR4     & 9.73 &  Order information is kept to meet legal requirements. \textbf{Your} personal
Information will be deleted  if you do not use this website for a month
                  \\ 

GDPR5     & 11.40 & If \textbf{you} have any questions or comments about this privacy policy, or the data collected, email ...                   \\  \rowcolor{Gray} 

GDPR6(d)     & 11.67 & 
\textbf{Your} data is stored safely and securely.
If we do lose \textbf{your} data we will be fined by the Information Commissioner  \\
\end{tabular}
\caption{Template text Gunning Fog Index}
\label{tab:TemplateGFI}
\end{table}

\begin{figure}[ht]
	\centering
       \includegraphics[width=\columnwidth]{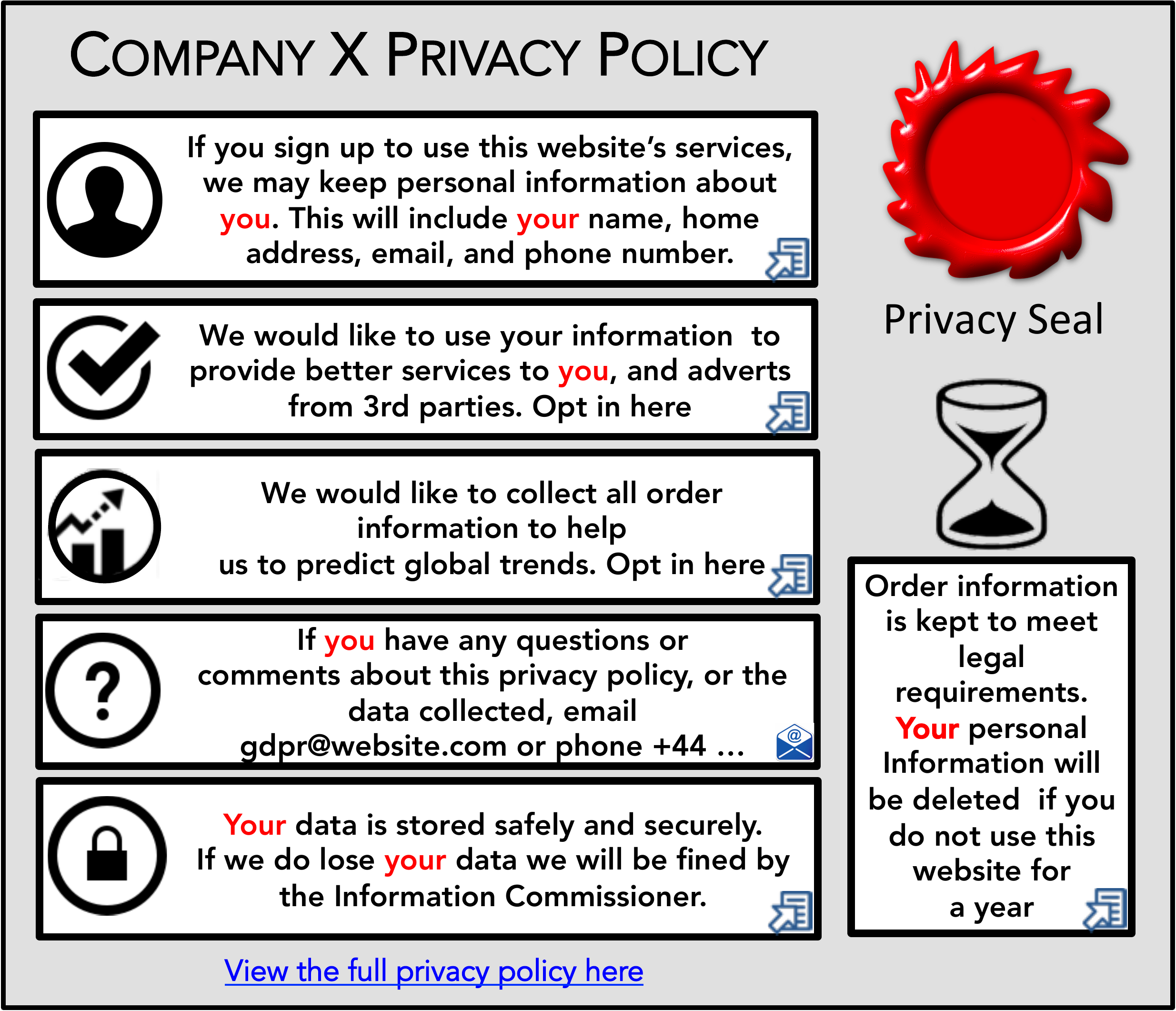}
	\caption{Usable GDPR-Compliant Privacy Policy Example}
	\label{fig:eg1}
\end{figure}

An exemplar GDPR-compliant and usable privacy policy was derived from the template shown in Figure \ref{fig:template} and is shown in Figure \ref{fig:eg1}. 
Company X, the company this privacy policy was tailored for, only uses their customers' information to detect global trends, and this is reflected in the middle box. This box, in particular, would reflect the purposes any particular organisation intends to use the customer's data for. The box on the right would also reflect a specific company's deletion policy; Company X only keeps data for 1 month --- others may keep it for 2 years. It is important that the actual policy is reflected here, so that the policy satisfies GDPR requirements. 

\section{Future Work}
\label{future}
The incoming GDPR legislation requires websites to obtain consent from their customers/users for any data collection to take place. This will inevitably lead to a veritable avalanche of consent requests as the GDPR deadline approaches. It is possible, as Schermer \etal \cite{schermer2014crisis} argue, that people will become desensitised  by all these requests and will start consenting without being fully aware of what they are consenting to. 
Adjerid \etal \cite{adjerid2013sleights} also argue that a myopic focus on transparency enhancement will not necessarily lead to improved and informed consent, especially when sites frame information differently. 
It would be very interesting to explore these apparent conundrums.

We proposed the use of a privacy seal to foster trust. A more detailed investigation is required in order to determine whether this is the most effective image to use. Some researchers found that privacy seals did enhance trust \cite{rifon2005your} but there is also evidence that users often misinterpret their message \cite{larose2006your}. 

\section{Conclusion}
\label{conc}

We publish this work  to provide guidance to designers and developers who need to incorporate privacy policies into their systems.  Our final template draws on the GDPR legislation and the research literature on usable design. 
We welcome feedback, particularly from those working in industry, to help us to refine and improve this template, to help it deliver maximum value.

\section*{Acknowledgements}
We thank Andrew Phillips for his feedback on an earlier draft of this paper. 

\balance
\bibliographystyle{IEEEtran}
\bibliography{refs}

\end{document}